\documentstyle[12pt]{article}    
\newcommand{\eq}{\begin{eqnarray}}
\newcommand{\en}{\end{eqnarray}}
\begin{document}

\hfill {\bf BUTP-99/24}

\vspace*{.3cm}
\begin{center}
{\Large\bf Hadronic Atoms in QCD}

\vspace*{.5cm}
J. Gasser\\
{\em Institute for Theoretical Physics, University of Bern,
Sidlerstrasse 5, CH-3012, Bern, Switzerland}\\
\vspace{0.3cm}
V. E. Lyubovitskij\footnote{Present address: 
Institute of Theoretical Physics, University of T\"{u}bingen, 
Auf der Morgenstelle 14, D-72076 T\"{u}bingen, Germany} \\
{\em Bogoliubov Laboratory of Theoretical Physics, Joint Institute
for Nuclear Research, 141980 Dubna, Russia and 
Department of Physics, Tomsk State University, 
634050 Tomsk, Russia}\\
\vspace{0.3cm}
A. Rusetsky\\
{\em Institute for Theoretical Physics, University of Bern,
Sidlerstrasse 5, CH-3012, Bern, Switzerland, 
Bogoliubov Laboratory of Theoretical Physics, Joint Institute
for Nuclear Research, 141980 Dubna, Russia and 
HEPI, Tbilisi State University, 380086 Tbilisi, Georgia}
\end{center}

\begin{abstract}
\setlength{\baselineskip}{2.6ex}
We propose a non relativistic effective Lagrangian approach to study
hadronic atom observables in the framework of QCD (including  photons). 
We apply our formalism to derive a general expression for the width of the 
$\pi^+\pi^-$ atom decaying into two neutral pions. It contains all terms at 
leading and next-to-leading order in isospin breaking. The result allows one 
to evaluate the combination $a_0-a_2$ of $\pi\pi$  $S$-wave scattering 
lengths from $\pi^+\pi^-$ lifetime measurements, like the one presently 
performed by the DIRAC experiment at CERN.
\end{abstract}

\setlength{\baselineskip}{2.6ex}

The DIRAC collaboration at CERN \cite{DIRAC} aims to measure the 
lifetime of the $\pi^+\pi^-$ atom (pionium) in its ground state at the 
10\% level. This atom decays predominantly into two neutral pions,
$\Gamma=\Gamma_{2\pi^0}+\Gamma_{2\gamma}+\ldots,\;$ with 
$\Gamma_{2\gamma}/\Gamma_{2\pi^0}\sim 4\cdot 10^{-3}$ \cite{DIRAC}.
The measurement of $\Gamma_{2\pi^0}$ allows one 
\cite{Bern}-\cite{Bilenky} to determine the difference $a_0-a_2$ 
of the strong $S$-wave $\pi\pi$ scattering lengths with isospin $I=0,2$. 
One may then confront the  predictions for this quantity obtained in 
standard ChPT \cite{ChPTlit,ChPT} with the lifetime measurement, and  
furthermore analyze the nature of spontaneous chiral symmetry breaking in 
QCD \cite{Stern}. In order to perform these investigations, one needs to 
know the theoretical expression for the width of pionium with a precision 
that properly matches the accuracy of the lifetime measurement of DIRAC. 
It is the aim of the present talk to derive a general formula \cite{Bern}
for the $\pi^+\pi^-$ atom decay width in the framework of QCD (including 
photons) by use of effective field theory techniques. Our result contains 
all terms at leading and next-to-leading order in the isospin breaking 
parameters $\alpha\simeq 1/137$ and $(m_u-m_d)^2$.
On the other hand, we expect that the contributions from 
next-to-next-to-leading order are completely negligible, at least for the
analysis of DIRAC data, and we therefore discard them here.

In several recent publications \cite{Labelle_Kong_Holstein}, the decay of 
$\pi^+\pi^-$ atoms has been  studied in the framework of a non relativistic 
effective Lagrangian approach - a method originally proposed by Caswell and 
Lepage \cite{Lepage} to investigate bound states in general. This method has 
proven to be far more efficient for the treatment of loosely bound 
systems - such as the $\pi^+\pi^-$ atom - than conventional approaches 
based on relativistic bound-state equations. It allows one e.g. to go beyond
the local approximation used in \cite{Sazdjian,Atom}. On the other hand, we 
are not aware of a systematic investigation of the decay of the $\pi^+\pi^-$ 
atom in this framework. In particular, the chiral expansion of the width has 
not been discussed, and a comparison of the corrections found in this 
framework with the results of \cite{Sazdjian,Atom} has never been provided. 
The main purpose of my talk - based on the papers \cite{Bern,Detail} - 
is to fill this gap.

The formation of the atom and its subsequent decay into two neutral 
pions is induced by isospin breaking effects in the underlying theory. In
the present framework, these are the electromagnetic 
interactions and  the mass difference of the up and down quarks. In the 
following, it is useful to count $\alpha\simeq 1/137$ and $(m_d-m_u)^2$ as 
small parameters of order $\delta$. 
More than forty years ago, Deser {\it et al.} \cite{Deser} derived the 
formula for 
the width of the $\pi^- p$ atom at leading order  in isospin symmetry 
breaking. Later in Refs. \cite{Uretsky,Bilenky}, this result was adapted to  
the $\pi^+\pi^-$ atom\footnote{There are a few misprints in 
Eq.~(6) for the pionium decay rate in Ref. \cite{Uretsky}. The correct 
result is displayed in Ref. \cite{Bilenky}.}. 
In particular, it was shown that - again at leading order in isospin symmetry 
breaking effects - the width  $\Gamma_{2\pi^0}^{\rm LO}$ of pionium is 
proportional to the square of the difference $a_0-a_2$,
\eq\label{deser}
\Gamma_{2\pi^0}^{\rm LO}=\frac{2}{9}\,\alpha^3 p^\star (a_0-a_2)^2\,\,   ;
\hspace*{1cm}
p^\star=(M_{\pi^+}^2-M_{\pi^0}^2-
\frac{1}{4}M_{\pi^+}^2\alpha^2)^{1/2}.   
\en
At leading order in $\delta$, the momentum $p^\star$ becomes
$\sqrt{2M_{\pi^+}(M_{\pi^+}-M_{\pi^0})}$ - this is the expression
used in \cite{Uretsky,Bilenky}. We prefer to use Eq.~(\ref{deser}),
because in this manner, one disentangles the kinematical corrections - 
due to the expansion of the square root - from  true dynamical ones.  
In our recent article \cite{Bern}, we derived a general expression 
for the pionium lifetime, that is valid  at leading and 
next-to-leading order in isospin breaking,
\eq\label{our}
\Gamma_{2\pi^0}=\frac{2}{9}\,\alpha^3 p^\star {\cal A}^2 (1+K). 
\en
The quantities ${\cal A}$ and $K$ are expanded in powers of
$\delta$. In particular, it has been shown in \cite{Bern} that\footnote{
We use throughout the Landau symbols $O(x)$ [$o(x)$] for quantities that
vanish like $x$ [faster than $x$] when $x$ tends to zero. Furthermore,
it is understood that this holds modulo logarithmic terms, i.e. we write also
$O(x)$ for $x\ln x$. 
}
\eq\label{eq21}
{\cal A}&=&-\frac{3}{32\pi}\,{\rm Re}A^{+-00}_{\rm thr}+o(\delta),
\en
where ${\rm Re}A^{+-00}_{\rm thr}$ is calculated as follows.
One evaluates  the relativistic scattering amplitude for the  
process $\pi^+\pi^-\rightarrow\pi^0\pi^0$ at order $\delta$ near 
threshold and removes the (divergent) Coulomb phase. The real part of this 
matrix element 
develops singularities that behave like $|{\bf p}|^{-1}$ and 
$\ln 2|{\bf p}|/M_{\pi^+}$ near threshold ($\bf{p}$ denotes the center of 
mass momentum of the charged pions). The 
remainder,  evaluated at the $\pi^+\pi^-$ threshold ${\bf p=0}$, equals
${\rm Re}A^{+-00}_{\rm thr}$. It contains terms of order $\delta^0$ and
$\delta$ and is normalized such that, in the isospin symmetry limit, 
${\cal A}=a_0-a_2$. Finally, the quantity $K$ starts at order 
$\alpha \ln\alpha$ - its explicit expression up to and including terms of
order $\delta$ is given by
\eq\label{K}
K=\frac{\Delta_\pi}{9M^2_{\pi^+}}\,(a_0+2a_2)^2
-\frac{2\alpha}{3}\,(\ln\alpha-1)\,(2a_0+a_2)+o(\delta), 
\hspace*{1cm}\Delta_\pi=M^2_{\pi^+}-M^2_{\pi^0}.
\en
In the derivation of Eqs.~(\ref{our}) - (\ref{K}), the 
chiral expansion has not been used. It is the aim of my talk to 
 show how this result can be derived
(details will be provided in a forthcoming publication \cite{Detail}). 

We proceed as follows. First, we display the non relativistic effective 
Lagrangian for pions, as derived from ChPT. Next, we formulate  resonance 
two-channel $\pi\pi$ scattering theory by applying Feshbach's 
projection technique \cite{Feshbach}. This method allows one to explicitly 
reveal the pole structure of the scattering matrix element and to obtain the 
equation for the bound-state energy of the $\pi^+\pi^-$ atom. 
Solving this equation, bound state characteristic of the hadronic 
atoms are given in terms of the couplings in the non relativistic Lagrangian. 
At the final stage, unknown couplings in the non relativistic Lagrangian can 
be expressed in terms of relativistic $\pi\pi$ scattering 
amplitudes through the matching procedure. Finally, as an illustration of our 
method, we derive Eqs.~(\ref{our}) - (\ref{K}).
 
The non relativistic effective Lagrangian 
${\cal L}={\cal L}_0 + {\cal L}_D + {\cal L}_C + {\cal L}_S$ - at the order 
of accuracy we are working here - consists of the free Lagrangian for  
charged and neutral pions (${\cal L}_0$), the "disconnected" piece 
(${\cal L}_D$) - providing  the correct relativistic relation between the 
energies and momenta of the pions - the Coulomb interaction piece 
(${\cal L}_C$), and the "connected" piece (${\cal L}_S$) which contains  
local four-pion interaction vertices:

\eq\label{Lagr_full}
{\cal L}_0&=& \sum_{i=\pm, 0}\,\pi_i^\dagger\biggl( i\partial_t-M_{\pi_i}+
\frac{\triangle}{2M_{\pi_i}}\biggr)\pi_i,\\[2mm]
{\cal L}_D&=&\sum_{i=\pm, 0}\,\pi_i^\dagger
\biggl(\frac{\triangle^2}{8M_{\pi_i}^3}+\cdots\biggr)\pi_i,\,\,\,\,\, 
{\cal L}_C\,\,=\,\,-4\pi\alpha(\pi_-^\dagger\pi_-)
\triangle^{-1}(\pi_+^\dagger\pi_+)+\cdots ,\nonumber\\[2mm]
{\cal L}_S&=&c_1\pi_+^\dagger\pi_-^\dagger\pi_+\pi_-
+c_2[\pi_+^\dagger\pi_-^\dagger(\pi_0)^2+{\rm h.c.}]
+c_3\,(\pi_0^\dagger\pi_0)^2\nonumber\\[2mm]
&+&c_4[\pi_+^\dagger\stackrel{\leftrightarrow}
{\triangle}\pi_-^\dagger(\pi_0)^2+\pi_+^\dagger\pi_-^\dagger\pi_0
\stackrel{\leftrightarrow}{\triangle}\pi_0+{\rm h.c.}]+\cdots,\nonumber
\en 
where $u\stackrel{\leftrightarrow}{\triangle}v\equiv 
u \triangle v + v \triangle u$. The coupling constants $c_i$ 
 are real at $O(\alpha)$ and are 
determined through matching to the relativistic theory. 

We now formulate the two-channel $\pi\pi$ scattering theory. We denote the
full Hamiltonian derived from (\ref{Lagr_full}) by $H=H_0+H_C+V$, with
$V=H_D+H_S$. The scattering operator $T$  obeys the Lippmann-Schwinger  
equation $T(z)=(H_C+V)+(H_C+V)G_0(z)T(z)$. The free and the Coulomb Green 
operators are defined as $G_0(z)=(z-H_0)^{-1}$ and $G(z)=(z-H_0-H_C)^{-1}$, 
respectively. The pole structure of the $T$-matrix is predominantly 
determined by the static Coulomb interaction $H_C$, whereas $V$ generates 
a small shift of the pole positions into the complex $z$-plane and will be 
treated perturbatively. To this end, we  use the method developed by 
Feshbach \cite{Feshbach} a long time ago. The $T$-matrix in our theory 
describes the transitions between  charged $|{\bf P},{\bf p}\rangle_+=
a^\dagger_+({\bf p}_1)a^\dagger_-({\bf p}_2)|0\rangle$ and neutral
$|{\bf P},{\bf p}\rangle_0=
a^\dagger_0({\bf p}_1)a^\dagger_0({\bf p}_2)|0\rangle$
states, where $a^\dagger_i$ denote the creation operators for 
non relativistic pions. Further,   ${\bf P}={\bf p}_1+{\bf p}_2$
and ${\bf p}=\frac{1}{2}({\bf p}_1-{\bf p}_2)$ are the CM 
and relative momenta of pion pairs, respectively. We work in the
CM system and remove the CM momentum  
from the matrix elements of any operator $R$, introducing the notation 
$_A\langle{\bf P},{\bf q}|R(z)|{\bf 0},{\bf p}\rangle_B=
(2\pi)^3\delta^3({\bf P})({\bf q}|r_{AB}(z)|{\bf p})$, 
where $A,~B=+,~0$. The operators $r_{AB}(z)$ act in the Hilbert space
of vectors $|{\bf p})$, where  the scalar product is defined as the integral
over the relative three-momenta of pion pairs. 

In order to avoid the complications associated with charged particles
in the final states, we consider the elastic  scattering process 
$\pi^0\pi^0\rightarrow \pi^0\pi^0$. In the vicinity of the $\pi^+\pi^-$
threshold, the scattering  matrix element develops a pole at \cite{Feshbach}
\eq\label{main}
z-E_0-(\Psi_0|\tau_{++}(z)|\Psi_0)=0,
\en
where $({\bf p}|\Psi_0)=\Psi_0({\bf p})$ stands for the unperturbed Coulomb 
ground-state wave function, and  $E_0$ is the corresponding ground-state 
energy. According to the conventional definition, the decay width is  
$\Gamma=-2{\rm Im} z$. The operator $\tau_{AB}(z)$ denotes the "Coulomb-pole 
removed" transition operator that satisfies the equation 
\eq\label{coupledchannel}
& &\tau_{AB}(z)=v_{AB}+v_{A+}\hat g_{++}(z) \tau_{+B}(z)
+ \frac{1}{4}v_{A0} g_{00}(z) \tau_{0B}(z),\nonumber\\[2mm]
& &({\bf q}|\hat g_{++}(z)|{\bf p}) = ({\bf q}|g_{++}(z)|{\bf p})  
-\frac{\Psi_0({\bf q})\Psi_0({\bf p})}{(z-E_0)}\,.
\en
It remains to solve Eq.~(\ref{main}) in the dimensional regularization scheme 
and with the use of the effective potential technique (see details in 
\cite{Bern,Detail}). 
We find that the width of the $\pi^+\pi^-$ atom - in terms of the effective
couplings $c_i$ -  is given by 
\eq\label{decaywidth}
\Gamma_{2\pi^0}=\frac{\alpha^3 M_{\pi^+}^3}{8\pi^2}\rho^{1/2}M_{\pi^0}
\biggl(1\hspace{-.8mm}+\hspace{-.8mm}\frac{5\rho}{8M_{\pi^0}^2}\biggr)
(c_2\hspace{-.5mm}-\hspace{-.5mm}2\rho c_4)^2\biggl(1-\rho c_3^2\frac{M_{\pi^0}^2}{4\pi^2}\biggr)
\biggl(1 - \frac{\alpha M_{\pi^+}^2}{4\pi}\xi c_1\biggr)+\cdots,
\en
where 
$\rho=2M_{\pi^0}(M_{\pi^+}-M_{\pi^0}-M_{\pi^+}\alpha^2/8)$,  
$\xi=2\ln\alpha-3+\Lambda+\ln(M_{\pi^+}^2/\mu^2)$,\\  
$\Lambda=(\mu^2)^{d-3}[(d-3)^{-1}-\Gamma'(1) - \ln4\pi]$. The ellipsis denotes
higher order terms in isospin breaking.
The divergent term proportional to $\Lambda$ 
stems  from a charged pion loop with one Coulomb photon exchange. 
It is removed by  the renormalization procedure in the scattering
sector. 
Next, we consider the matching procedure, which relates the effective 
couplings $c_i$  to the $\pi\pi$ scattering amplitudes for three channels 
$\pi^+\pi^-\to\pi^0\pi^0$,  $\pi^+\pi^-\to\pi^+\pi^-$ and 
$\pi^0\pi^0\to\pi^0\pi^0$ evaluated in the relativistic theory \cite{Knecht}: 
\eq\label{matching1}
3M_{\pi^+}^2\, c_1=4\pi\, (2a_0+a_2) +o(\delta), \hspace*{1cm} 
3M_{\pi^+}^2\, c_3=2\pi\, (a_0+2a_2) +o(\delta), 
\en
\eq\label{matching2}
{\cal A}=-\frac{3}{8\pi}M_{\pi^+}^2\biggl[2c_2-
4\Delta_\pi\biggl(c_4+\frac{c_2c_3^2}{8\pi^2}M_{\pi^0}^2\biggr)
+\frac{\alpha M_{\pi^+}^2}{4\pi}\biggl(1-\Lambda-\ln\frac{M_{\pi^+}^2}
{\mu^2}\biggr)c_1c_2\biggr] + o(\delta) 
\en
Substituting Eqs.~(\ref{matching1}), (\ref{matching2}) into 
Eq.~(\ref{decaywidth}), we finally arrive at the general formula for the
$\pi^+\pi^-$ atom decay width in QCD, given by
Eqs.~(\ref{our}) - (\ref{K}).
It is our opinion that these equations
finalize the attempts to calculate the width $\Gamma_{2\pi^0}$ at 
next-to-leading order, relegating the problem  to the evaluation of the 
physical on-mass-shell scattering amplitude for the process 
$\pi^+\pi^-\rightarrow\pi^0\pi^0$ to any desired order 
in the chiral expansion. Numerical analysis 
of the $\pi^+\pi^-$ atom lifetime in ChPT at one loop, including a comparison 
with recent work in literature, was performed recently 
in Ref. \cite{Numerics}. 

In conclusion, we have evaluated the width $\Gamma_{2\pi^0}$ of the
$\pi^+\pi^-$ atom in its ground state at leading and next-to-leading
order in isospin breaking. The non relativistic effective 
Lagrangian approach of Caswell and Lepage \cite{Lepage} appears to be
an extremely suitable tool for this purpose, that allows one 
to completely solve this problem. Its usefulness may be seen even more clearly
for the case of $p\pi^-$, $pK^-$, $d\pi^-$, $dK^-$ atoms, studied in ongoing 
or planned  experiments (PSI, KEK, DA$\Phi$NE), because this approach 
 trivializes the spin-dependent part of the problem. 

{\it Acknowledgments}. 
V. E. L. thanks the Organizing Committee of MENU99 Symposium for financial 
support and University of Bern for hospitality where this work was done. 
The work was supported in part by the Swiss National Science
Foundation, and by TMR, BBW-Contract No. 97.0131  and  EC-Contract
No. ERBFMRX-CT980169 (EURODA$\Phi$NE).

\end{document}